\documentclass{ws-ijmpe}

\usepackage{dcolumn,array,bm}

\usepackage{graphicx}

\newcommand{\beq}{\begin{equation}}
\newcommand{\eeq}{\end{equation}}
\newcommand{\ba}{\begin{array}}
\newcommand{\ea}{\end{array}}
\newcommand{\bn}{\begin{eqnarray}}
\newcommand{\en}{\end{eqnarray}}

\begin{document}

\markboth{S.G. Rohozi\'nski et al.}{
Signatures of chirality}

\catchline{}{}{}{}{}

\title{SIGNATURES OF CHIRALITY IN THE CORE-PARTICLE-HOLE SYSTEMS
}

\author{\footnotesize STANIS{\L}AW G. ROHOZI\'NSKI
}

\address{Institute of Theoretical Physics, University of Warsaw, ul. Ho\.za
69\\
00-681 Warsaw, Poland
\\
Stanislaw-G.Rohozinski@fuw.edu.pl}

\author{\footnotesize LESZEK PR\'OCHNIAK}

\address{
Institute of Physics,
UMCS,
Pl. M. Curie-Sk{\l}odowskiej 1 \\
20-031, Lublin, Poland 
}

\author{\footnotesize CHRYSTIAN DROSTE}
\address
{
Institute of Experimental Physics, Faculty of Physics, University of Warsaw, ul. Ho\.za 69\\
 00-681, Warsaw, Poland
}

\author{\footnotesize KRZYSZTOF STAROSTA}

\address
{Department of Chemistry, Simon Fraser University, 8888 University Drive \\
Burnaby, British Columbia, V5A1S6, Canada 
}

\maketitle

\begin{history}
\received{(received date)}
\revised{(revised date)}
\end{history}
\begin{abstract}
An odd-odd nucleus is treated as the core-particle-hole system. The core is described by the Bohr Hamiltonian.
Different collective potentials of the core are investigated. The odd particle and hole are assumed to be in the symmetric
$\pi h_{11/2}\otimes\nu h_{11/2}^{-1}$ configuration. 
Signatures of chirality in the odd-odd nucleus spectra are observed. The sufficient condition for the appearance
of signatures of chirality in the core-particle-hole system is the $\alpha$-symmetry of the core provided the particle-hole configuration
of the odd valence particles is symmetric.
\end{abstract}

\section{Introduction}\label{intro}
Chirality in nuclei is for the decade
a hot topic  in the nuclear structure physics
of the odd and odd-odd nuclei. It allows us to interpret
the spectra of some odd-odd nuclei in a simple way.
The nuclear chiral system is the one represented by three noncoplanar angular momentum (pseudo)vectors.
The three vectors can have one of the two possible handednesses or chiralities. 
 
The original chiral system modelling an odd-odd nucleus consisted of the triaxial rigid rotor, the odd nucleon (say, the proton) and 
the odd nucleon-hole (neutron-hole) occupying the same $j$-shell in a deformed mean field created by the core\cite{Fra97,Koi04}. 
The system has its natural intrinsic frame of 
reference: the three principal axes of the rotor. In the quasi-classical picture the proton and the neutron-hole circulates 
around the short and the long axis, recpectively, whereas the rotor rotates around its intermediate axis. Then, the angular momenta of the three are directed along the principal axes and form either the left-handed or the right-handed system depending on the directions of circulation. 
The quantum picture is a bit more complicated because the angular momenta of the proton, the neutron-hole and the core are not conserved 
and  mean directions of them can be determined at most. It would be interesting to construct explicitly the left- and right-handed quantum 
states of the system.

In our approach to the problem of chirality we use a much more involved and realistic model to describe odd-odd nuclei. We couple the proton and the neutron-hole
to any core of our choice through the quadrupole-quadrupole forces and we allow the valence particle and hole to occupy various orbitals.
When the core is realistic andthe single-particle basis is sufficiently large, and the pairing forces are not important the model is able to describe
a realistic odd-odd nucleus. The model is called the Core-Particle-Hole Coupling (CPHC) model\cite{Sta02}. It is a special version of 
the Core-Quasiparticle-Pair Coupling model\cite{Kle04,Roh05} in which the pairing interaction is neglected. There is no need to introduce an intrinsic frame of reference for the valence particles because the explicit interaction of them with the core is taken instead of the strong coupling scheme.
The chirality phenomenon can be investigated by observing signatures of chirality
in the excitation spectrum of the odd-odd nucleus.
 
The main signatures of the chirality in nuclei are:
  \begin{enumerate}
  \item 
Appearence of a pair of almost degenerate $\Delta I=1$ bands of the same parity
called chiral partner bands.
\item Similar electromagnetic moments and transition probabilities in the corresponding states of chiral bands.
    \item
  Staggering of the intraband and interband M1 and the $\Delta I=1$ E2 transitions.
\end{enumerate}

In the present study we take simple models of the cores just to connect the appearance of the chirality signatures with the characteristic features  
of the nucleus with a particular core. However, we do not investigate an effect of different configurations of the valence particle and hole 
which is also important for the chirality phenomenon. Here we consider only the case of the particle and hole both occupying the same $j$-shell.
A more comprehensive analysis will be published elsewhere\cite{Dro10}, which will comprise in particular the problem of asymmetric configurations of the particle and hole. Below we present only some examples of the results of our investigation.

In Sec. ~\ref{model}, we present briefly grounds of the calculations.
Sec.~\ref{symm} contains the discussion of  properties of the spectra of odd-odd nuclei with  the $\alpha$-symmetric cores of different rigidities 
from the point of view of the manifestation of chirality. A similar discussion for the case of the $\alpha$-asymmetric cores is included in Sec.~\ref{asymm}.
Sec.~\ref{sum} makes the summary of our study.

\section{Core-Particle-Hole Coupling Model}\label{model}
The odd-odd nucleus is treated as the three-body system: the even-even core, the proton, and the neutron-hole. It means that the
states of the odd-odd nucleus with the proton and neutron numbers $Z$ and $N$, respectively, are assumed in the following form:
\begin{eqnarray}
&& |Z,N;iIM\rangle   \nonumber \\
 &&=\sum_{\rho ,\sigma}\sum_{L,R,r}U_{Ii}(\rho ,\sigma ,L,R ,r)  
 \left[\left[a^{\dag}_{\pi\rho}\times\tilde{a}_{\nu\sigma}\right]_L\times |Z-1,N+1;rR\rangle\right]_{IM}
\label{states} 
\end{eqnarray}
where $a^{\dag}_{\pi\rho}$ and $\tilde{a}_{\nu\sigma}$ are the proton and the neutron-hole creation operator in the single-particle
states $\rho =n_{\rho}l_{j_{\rho}}$ and $\sigma =n_{\sigma}l_{j_{\sigma}}$, respectively, and $ |Z-1,N+1;rRM_R\rangle$ is the even-even
core state with the angular momentum quantum numbers $RM_R$ and the remaining numbers $r$.
The quadrupole-quadrupole two-body interaction between the proton, the neutron and the core is assumed.
The coupling constant $\chi_2= 40$MeV/b$^2$, relatively strong, is taken in the present calculations.
The proton and neutron numbers are put $Z=57,\  N=71$. It would correspond to $^{128}$La. However, we use different fictitious cores 
and, in fact, we consider a fictitious nucleus.
Details of the model and the principles of calculations are presented in \cite{Sta02,Dro09}.

\subsection{Description of the core}\label{core}
The core states $|Z-1,N+1;rRM_R\rangle$ are described by a version of the Bohr Hamiltonian (cf. e.g. \cite{Pro09}) in the following form:
\begin{eqnarray}
&&H(\beta ,\gamma ,\Omega )  
=-\frac{1}{2B_{\beta\beta}}\frac{1}{\beta^4}\frac{\partial}{\partial\beta}\left(\beta^4\frac{\partial}{\partial\beta}\right) \nonumber \\
&&-\frac{1}{2B}\frac{1}{\beta^2\sin{3\gamma}}\frac{\partial}{\partial\gamma}\left(\sin{3\gamma}\frac{\partial}{\partial\gamma}\right)
-\sum_{k=1}^3\frac{R_k^2(\Omega )}{\sin^2{(\gamma -2\pi k/3)}} \nonumber \\
&& +{\frac{1}{2}}V_C\beta^2+\left(G+h_1\cos{3\gamma}+h_2(\cos^2{3\gamma}-1)^\kappa\right) 
\left(\exp{(-\beta^2/d^2)}-1\right) \label{Bohr}
\end{eqnarray}
where variables $\beta$ and $\gamma$ are the Bohr
deformation parameters, $\Omega$ stands for the three Euler angles of orientation of the body-fixed system and  
$R_k(\Omega )$ for $k=1,\ 2,\ 3$ 
are the three (dimensionless) intrinsic components of angular momentum.
The rigid rotor model\cite{DF58} is an extreme case of the Bohr Hamiltonian for the infinite stiffnesses of the collective potential against $\beta$ and $\gamma$.

In the calculations the parameters $h_1$, $h_2$ and $\kappa$ are varied in order to have a definite dependence 
of the collective potential on $\gamma$. The vibrational inertial parameter is settled equal to $B_{\beta\beta}=250$/MeV.
The remaining parameters of the Hamiltonian
$B$, $V_C$, $G$ and $d$ are selected in such a way 
that the equilibrium deformation is about $\beta =0.25$ and the values of energy of the lowest excited state $E(2^+_1)$ and reduced transition probability 
$B(\mathrm{E2};2^+_1\rightarrow 0^+_1)$ are always close to  values $E(2^+_1)=354$ keV and  
$B(\mathrm{E2};2^+_1\rightarrow 0^+_1)=0.282\ e^2\mathrm{b}^2$ close to the experimental values for 
$^{128}_{\phantom{1}56}$Ba ($A=128$, $Z-1=56$). It is done so to have
the same scale of results.

The Bohr Hamiltonian is said to be the $\alpha$-symmetric if it is invariant under the O(5) inversion $\alpha_{2\mu}\to -\alpha_{2\mu}$ of
the laboratory  quadrupole variables related to $\beta ,\ \gamma ,\  \Omega$ as follows:
\begin{equation}
\alpha_{2\mu}(\beta ,\gamma ,\Omega ) 
=  D^2_{\mu 0}(\Omega )\beta\cos{\gamma}+\frac{1}{\sqrt{2}}\left(D^2_{\mu 2}(\Omega )
+D^2_{\mu -2}(\Omega )\right)\beta\sin{\gamma}\label{lab}
\end{equation}
Then the collective states possess the definite $\alpha$-parity $p_{\alpha}=\pm 1$ \cite{Bes59}. The O(5) inversion in terms of the intrinsic variables
is: 
\beq
(\beta ,\gamma ,\Omega )\to (\beta ,\pi /3-\gamma ,R_1(\pi /2)\Omega )
\label{intr}
\eeq
where $R_1(\pi /2)$ is the rotation by $\pi /2$ around the body-fixed 1-axis. 

\subsection{The particle-hole configuration}\label{conf} 
Should the single-particle bases for the proton and the neutron-hole contain the same set of orbitals and the single-particle energies of the proton and neutron are equal to each other the proton and neutron-hole subsystem is the proton-neutron symmetric  \cite{Koi04}
i.e. is invariant under interchange of the proton and the neutron-hole  occupying the same orbital 
$\pi\rho\rightleftharpoons\nu\rho^{-1}$.

In the present study we restrict ourselves to the one simple symmetric particle-hole configuration $\pi h_{11/2}\otimes\nu h_{11/2}^{-1}$.

\section{The $\alpha$-symmetric cores}\label{symm}
\begin{figure}[htb]
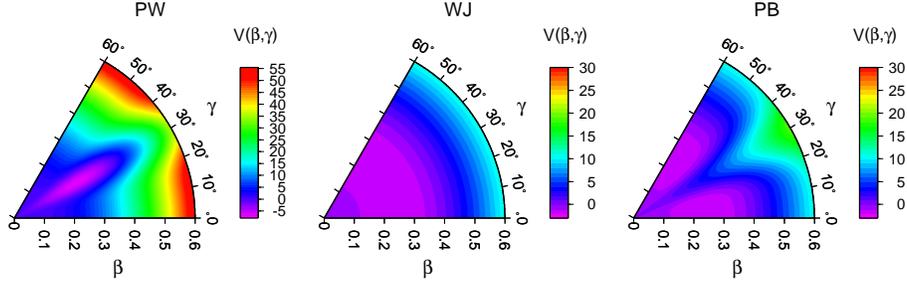

\psfig{file=sym_pot_pw.eps,width=3.8cm}\hspace{0.2cm}
\psfig{file=sym_pot_wj.eps,width=3.8cm}\hspace{0.2cm}
\psfig{file=sym_pot_pb.eps,width=3.8cm}
\caption{Contour maps of the three collective potentials in the Bohr Hamiltonian (\ref{Bohr}).  Left: the potential with a well in $\gamma$ (PW), ($h_1=0$, $h_2=20$MeV, $\kappa =4$). Middle: 
the $\gamma$-independent potential (WJ), ($h_1=h_2=0$). Right: the potential with a barrier in $\gamma$ (PB),
($h_1=0$, $h_2=-8$MeV, $\kappa =4$).}
\label{figsympot}
\end{figure}

In our earlier paper\cite{Dro09} we considered two models for the core:  the Davydov-Filippov (DF) model\cite{DF58} with parameter 
$\gamma =30^{\circ}$ (the maximal triaxiality), and the Wilets-Jean (WJ) model\cite{WJ56} which uses the $\gamma$-independent collective potential 
(the $\gamma$ softness or instability). The odd-odd nuclei in both cases of the core and a symmetric particle-hole configuration manifest the chirality 
signatures in their excitation spectra. The question arises 

how this is in the case of a finite rigidity or  an incomplete softness. 
In order 
to answer this question
we present here the results of calculations with the cores described by the three collective potentials shown 
in 
Fig. \ref{figsympot}. All the three potentials are the $\alpha$-symmetric. We look for signs of the chirality in the odd-odd nuclei with these three cores. 

\begin{figure}[htb]
\centerline{\psfig{file=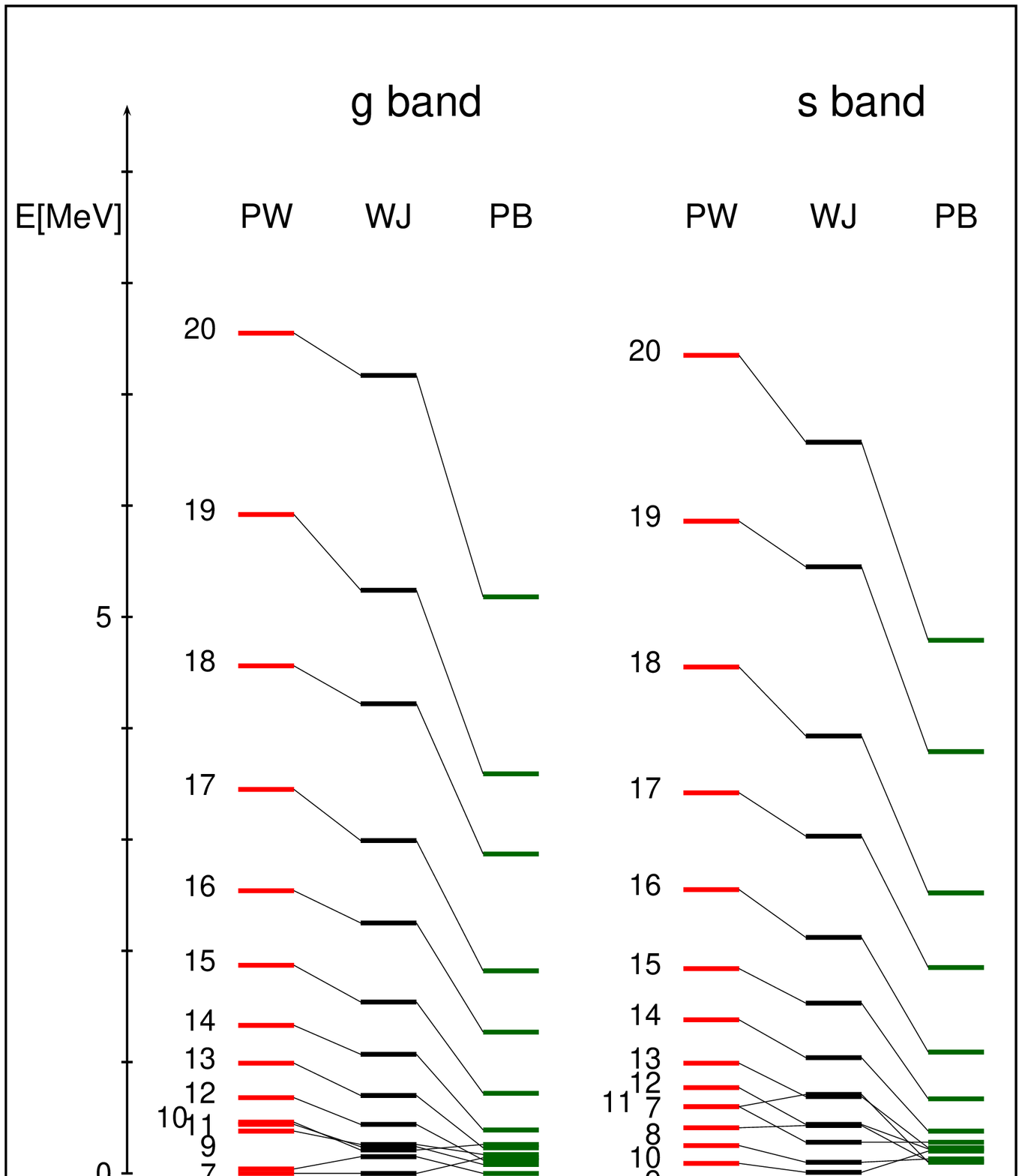,width=5.0cm}}
\caption{The energy levels $E(I_{\mathrm{b}})$ in the ground (b=g) and side (b=s) bands of the odd-odd nucleus calculated in the three versions of the core: with the potential well (PW),
with the $\gamma$-unstable potential (WJ) and with the potential barrier (PB).} 
\label{figsymband}
\end{figure}

The calculated energy levels $E(I_{\mathrm{b}})$ of the ground band (b=g) and the
side band (b=s)  in the three odd-odd nuclei in question are shown in Fig.~\ref{figsymband}. It is seen that both bands are more and more
stretched for bigger and bigger rigidity (the rigidity of the barrier can be treated as negative). The bands in the nucleus with the DF core
 would be still more stretched (cf. ref. \cite{Dro09}). However, in the all three cases the splitting between the states of the same spin $I$ in both bands 
is relatively small and thus the bands can be treated as the chiral partner bands. It turns out that the values of the magnetic dipole moments 
$\mu (I_{\mathrm{b}})$ do not depend
practically on the version of core and on the band b. The same can be said about the electric quadrupole moments which values are in all the cases  
close to zero (smaller than the corresponding single-particle estimation), cf. Fig. ~\ref{figasband} in Sec. ~\ref{asymm}. The values of the reduced transition probabilities of the intra-band stretched E2
transitions $B(\mathrm{E2}; I_{\mathrm{b}}\to (I-2)_{\mathrm{b}})$ depend indeed on the version of core but are close to each other in both bands b. 

\begin{figure}[htb]
\centerline{\psfig{file=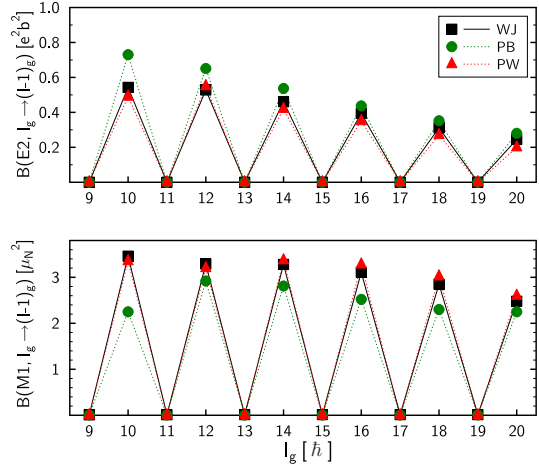,width=7cm}}

\caption{Reduced transition probabilities of the intra-band E2 and M1 transitions $B(\mathrm{E2}; I_{\mathrm{g}}\to (I-1)_{\mathrm{g}})$ (upper panel) 
and $B(\mathrm{M1}; I_{\mathrm{g}}\to (I-1)_{\mathrm{g}})$ (lower panel) calculated in the three cases of the core.}
\label{figsymbemgg}
\end{figure}

The reduced transition probabilities of the $\Delta I=1$ intra-band g$\to$g  both E2 and M1 transitions depend weakly on the
rigidity of the core and show very regular and strong staggering. We see it in 
Fig.~\ref{figsymbemgg}.
The inter-band s$\to$g transitions behave in very similar way to the intra-band ones. The only difference is that the staggering
of their values is in the opposite phase to that of the g$\to$g transitions.

\section{The $\alpha$-asymmetric cores}\label{asymm}
\begin{figure}[htb]
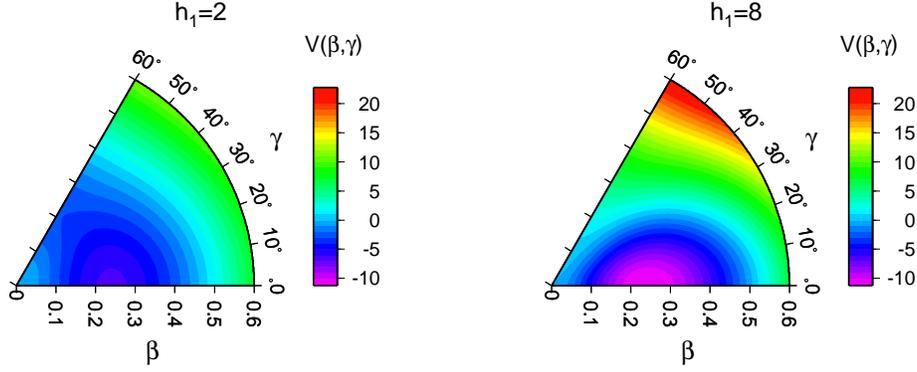

\psfig{file=as_pot_h2.eps,width=5cm}\hspace{2cm}
\psfig{file=as_pot_h8.eps,width=5cm}

\caption{Contour maps of the two $\alpha$-asymmetric collective potentials in the Bohr Hamiltonian (\ref{Bohr}).  Left: the potential with a weaker 
asymmetry ($h_1=2$MeV, $h_2=0$, $\langle\gamma\rangle =20.7^{\circ}$)  
Right: the potential with a stronger asymmetry
($h_1=8$MeV, $h_2=0$, $\langle\gamma\rangle =15.0^{\circ}$).}
\label{figaspot}
\end{figure}

\begin{figure}[htb]
\psfig{file=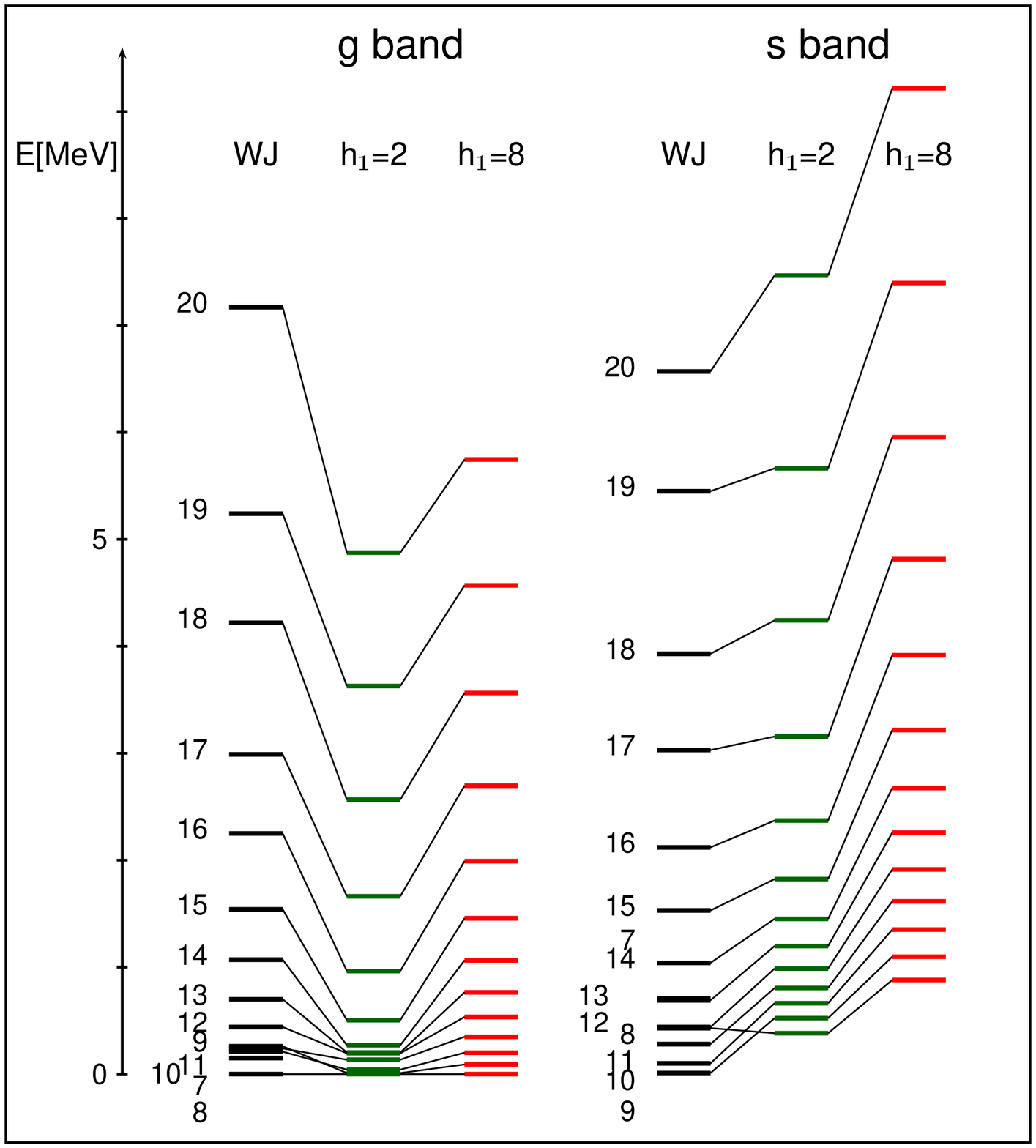,width=4.8cm}\hspace{0.5cm}
\psfig{file=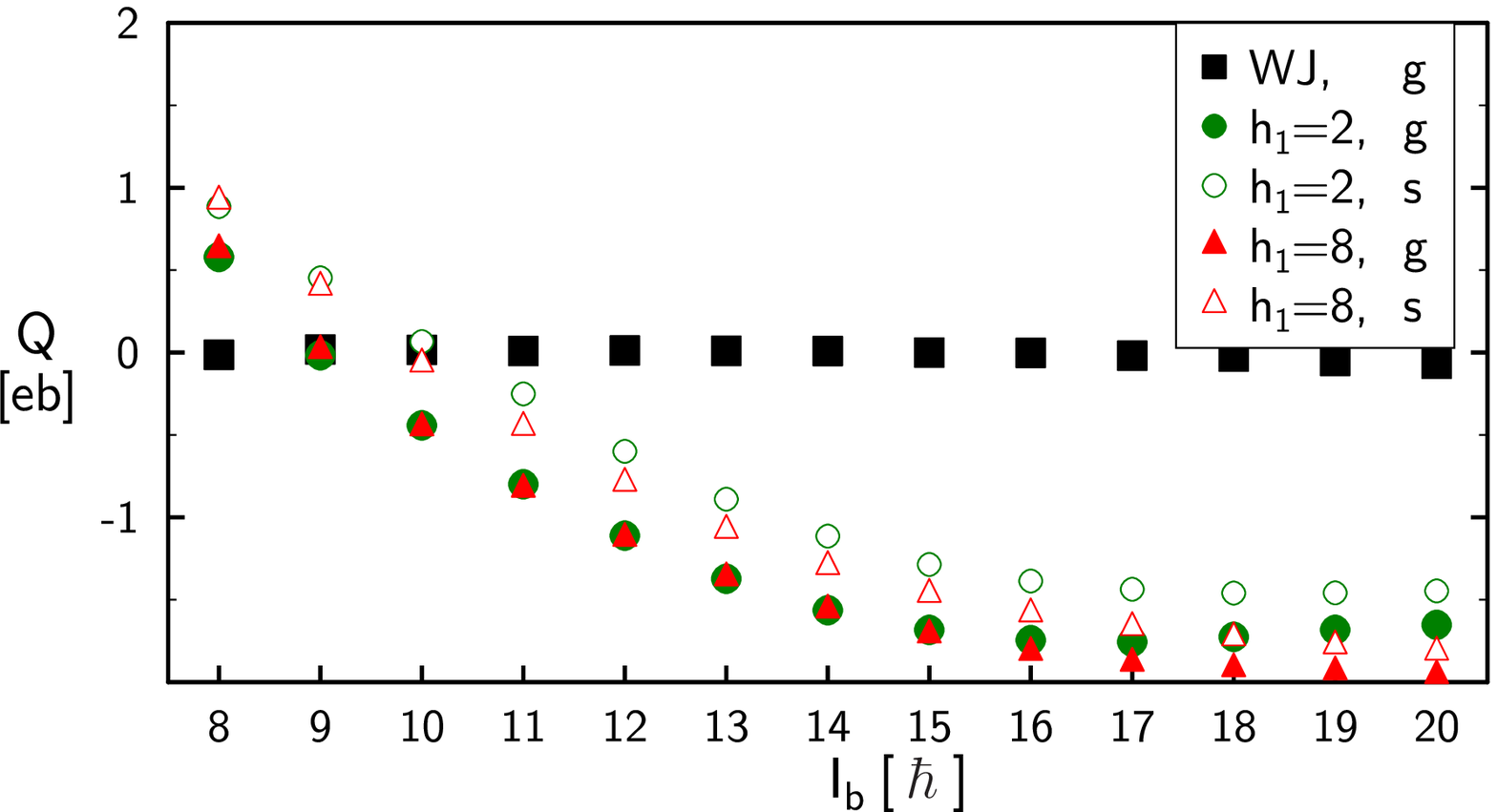,width=7cm}

\caption{The ground (b=g) and side (b=s) bands in the odd-odd nucleus calculated for the two cores with the potential parameter $h_1=2$MeV
and $h_1=8$MeV, respectively.
 Left panel: the energy levels $E(I_{\mathrm{b}})$. 
Right panel: the values of the electric quadrupole $Q(I_{\mathrm{b}})$.}
\label{figasband}
\end{figure}

\begin{figure}[htb]
\centerline{\psfig{file=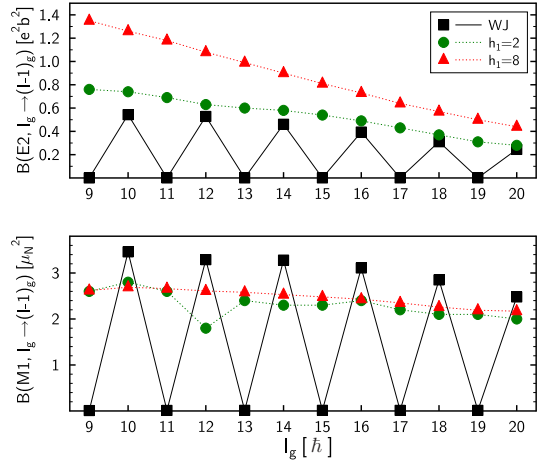,width=7cm}}

\caption{Reduced transition probabilities of the intra-band E2 and M1 transitions $B(\mathrm{E2}; I_{\mathrm{g}}\to (I-1)_{\mathrm{g}})$ (upper panel) 
and $B(\mathrm{M1}; I_{\mathrm{g}}\to (I-1)_{\mathrm{g}})$ (lower panel) calculated for the two $\alpha$-asymmetric cores:
the core with weaker asymmetry ($h_1=2$MeV) and the one with stronger asymmetry ($h_1=8$MeV).}
\label{figasbemgg}
\end{figure}

\begin{figure}[htb]
\centerline{\psfig{file=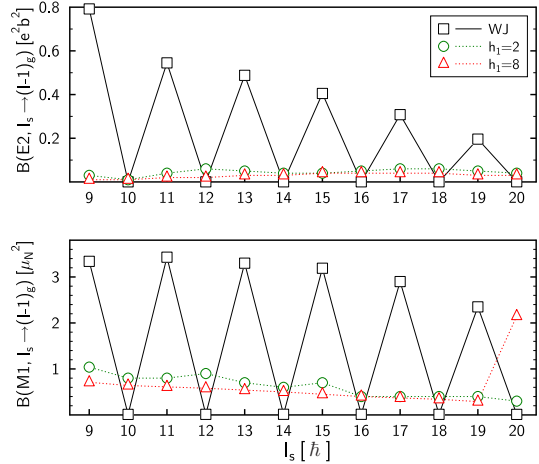,width=7cm}}

\caption{The same as in Fig. ~\ref{figasbemgg} but for the reduced transition probabilities of the inter-band E2 and M1 transitions $B(\mathrm{E2}; I_{\mathrm{s}}\to (I-1)_{\mathrm{g}})$ (upper panel) 
and $B(\mathrm{M1}; I_{\mathrm{s}}\to (I-1)_{\mathrm{g}})$ (lower panel).}
\label{figasbemsg}
\end{figure}

The rigid DF core with the maximal triaxiality ($\gamma =30^{\circ}$) is $\alpha$-symmetric. Even small deviation of $\gamma$ from the maximal
triaxiality causes that  the staggering of the $\Delta I=1$ transition probability values vanishes\cite{Dro09}. It is interesting whether a small
$\alpha$-asymmetry of a soft core gives a similar effect. To check the effect we perform the calculations for the odd-odd nuclei with 
the $\alpha$-asymmetric cores which have the ground state mean values of $\gamma$ around 21$^{\circ}$ ($h_1=2$MeV) and 15$^{\circ}$ 
($h_1=8$MeV), respectively. The corresponding collective potentials of the cores are shown in Fig. \ref{figaspot}.
It turns out that the ground and side bands loose their chiral-partner-band character when the asymmetry of the collective potential rises because  
the splitting of them rises too. 
It is seen in Fig. \ref{figasband}. The absolute values of the electric quadrupole moments are obviously bigger
than those in the case of the WJ core but they are still close to each other in both bands as is seen in Fig. \ref{figasband}. 
On the other hand, the magnetic dipole moments and the 
reduced transition probabilities of the intra-band stretched E2 transitions depend weakly on the case of the core and their values 
do not depart much from those for the WJ core. The reduced transition probabilities of the $\Delta I=1$
intra-band  E2 transitions increase together with the increase of asymmetry whereas those of the M1 ones depend weakly on the asymmetry.
The reduced probabilities of the inter-band transitions, both E2 and M1, decrease. The staggering 
of the values of the all transition probabilities vanishes immediately with a deviation from the $\alpha$-symmetry of the core. It is shown in 
Figs. \ref{figasbemgg} and \ref{figasbemsg}. Apparently, this phenomenon does not depend on rigidity of the core.

\section{Summary}\label{sum}
The signatures of chirality in the odd-odd nuclei treated as a three-body core-particle-hole systems have been observed from the laboratory frame of reference. Although the body-fixed frame has been used in the description of the core, it is not useful in the description of the three-body system
because the core characteristics enter the description of the odd-odd nucleus only through the energies
of the collective levels and the quadrupole matrix elements within the collective states.
No assumptions on the chiral geometry have been made.
It has been found that the sufficient condition for the core-particle-hole system to manifest the all signatures of chirality is the $\alpha$-symmetry of the core provided the particle-hole configuration is the proton-neutron symmetric.

\section*{Acknowledgements}

The authors would like to acknowledge help of A. Chester in preparation of the manuscript. This work was performed within the frame of project No. G-POOL/2009/0  and supported in part by the Polish Ministry of Science
under Contract No.~N~N202~328234.

\end{document}